\documentstyle[prd,aps,epsfig,floats]{revtex}
\begin{document}
\draft
\wideabs{
\title{RUGGED METROPOLIS SAMPLING WITH SIMULTANEOUS UPDATING OF TWO 
DYNAMICAL VARIABLES}

\author{ Bernd A. Berg$^{\rm \,a,b}$ and
         Huan-Xiang Zhou$^{\rm \,a,b,c}$ }

\address{ 
$^{\rm \,a)}$ Department of Physics, Florida State University,
  Tallahassee, FL 32306-4350\\
$^{\rm \,b)}$ School of Computational Science, Florida State 
  University, Tallahassee, FL 32306-4120\\
$^{\rm \,c)}$ Institute of Molecular Biophysics, Florida State 
  University, Tallahassee, FL 32306-4380\\
} 
% (E-mail: berg@csit.fsu.edu)\\ 

% \date{January 6, 2005}
\date{\today }

\maketitle
\begin{abstract}
The Rugged Metropolis (RM) algorithm is a biased updating scheme, 
which aims at directly hitting the most likely configurations in a 
rugged free energy landscape. Details of the one-variable (RM$_1$) 
implementation of this algorithm are presented. This is followed 
by an extension to simultaneous updating of two dynamical 
variables (RM$_2$). In a test with Met-Enkephalin in vacuum RM$_2$ 
improves conventional Metropolis simulations by a factor of about 
four. Correlations between three or more dihedral angles appear to
prevent larger improvements at low temperatures. We also 
investigate a multi-hit Metropolis scheme, which spends more CPU 
time on variables with large autocorrelation times.
\end{abstract}
\pacs{PACS: 05.10.Ln, 87.15-v, 87.14.Ee.}
}

\narrowtext

\section{Introduction} \label{sec_intro}

Simulations of biomolecules are one of the major challenges in 
computational science. Rugged free energy landscapes are typical
for such systems. In this context a Rugged Metropolis (RM) algorithm 
was introduced in Ref.~\cite{Be03}. The motivation of RM was an 
elaboration of the funnel picture of protein folding, which was 
originally formulated by Bryngelson and Wolynes~\cite{BrWo87}. 
RM uses a biased Metropolis 
algorithm, with the bias of the updating proposal obtained using 
data from previous simulations at higher temperatures. Although the 
possibility of constructing biased Metropolis algorithms has been 
known for many years \cite{Ha70}, and these have occasionally been 
used in the statistical physics \cite{Br85,MHB86}
and bio-chemical \cite{DeBa96,Fa01,Jo02} literature, it seems that a 
systematic understanding of the possibilities of biased Metropolis 
procedures is still in its infancy. For instance, it was only 
recently noted \cite{BaBe05} that biased one-variable updates 
allow one to imitate the heat bath (Gibbs sampler) updates and can 
still be efficient when the conventional calculation of heat bath 
probabilities becomes prohibitively slow.

The RM approach is distinct from generalized ensemble simulations. 
Generalized ensembles (for reviews and recent work see \cite{MSO01}) 
also use information from higher temperatures, but in an entirely 
different way. In a sense generalized ensembles build bridges in a 
rugged free energy landscape, while the RM scheme aims to enhance the 
likelihood for a direct hit of the needle in the haystack. In fact 
RM updates can be implemented within any generalized ensemble. In a 
test case of RM updates within a replica exchange simulation, the 
improvement was multiplicative~\cite{Be03}.
% That ``needle'' may have a high statistical probability in the 
% canonical ensemble, but nevertheless can not be found by the 
% conventional Metropolis algorithm due to free energy barriers.

The main technical challenge within the RM scheme is to obtain, from 
available time series data, estimates of the multi-variable probability 
densities (pds) in a form that allows for their fast numerical 
evaluation. So far this was only achieved for one-variable pds, 
resulting in the RM$_1$ update scheme. However, it is well-known that 
many degrees of freedom in a protein molecule are coupled. In addition,
one needs multi-variable moves to avoid steric clashes \cite{GoSc70} 
(cf. \cite{Jo02} and references therein for more recent literature). 
As a next approximation to the desired RM probabilities, in this paper 
we deal with pds of two variables to develop and test the corresponding 
RM$_2$ update scheme.

In the present paper all our Monte Carlo (MC) simulations are 
done in the canonical ensemble for the brain peptide Met-Enkephalin 
in vacuum, which has been a frequently used test case since its 
initial numerical investigation in Ref.~\cite{LiSh87}. For this
(artificial) system the coil-globule transition temperature is at
$T_{\theta}\approx 295\,$K and the folding temperature is at
$T_f\approx 230\,$K according to Ref.~\cite{HMO97}. Long living 
traps are found at the glass transition temperature, which is for
Met-Enkephalin below the folding temperature at 
$T_g\approx 180\,$K~\cite{AH00}.  In our 
simulations we cover a range from $400\,$K down to $220\,$K and
measure integrated autocorrelation times (see, e.g., Ref.~\cite{bbook} 
for the definition) to determine the performance of our algorithms. 

In section~\ref{sec_RM1} we review the RM scheme and its RM$_1$ 
approximation, filling in many details which inevitably had to be 
omitted in the letter format of Ref.~\cite{Be03}. On the fly we also 
investigate a multi-hit updating procedure, which spends more computer 
time on variables with large integrated autocorrelation times. In 
section~\ref{sec_RM2} we introduce and test a RM$_2$ scheme. Summary 
and conclusions follow in section~\ref{sec_summary}. 

\section{RM and the RM$_1$ approximation} \label{sec_RM1}

We consider biomolecule models for which the energy $E$ is a function 
of a number of dynamical variables $v_i,\, i=1,\dots,n$. The
fluctuations in the Gibbs canonical ensemble are described by a 
probability density~(pd) $\rho(v_1,\dots ,v_n; T)$, where $T$ is 
the temperature. To be definite, we use in the following the all-atom 
energy function ECEPP/2 (Empirical Conformational Energy Program for 
Peptides)~\cite{SNS84}. Our dynamical variables $v_i$ are the dihedral 
angles, each chosen to be in the range $-\pi\le v_i< \pi$, so that the 
volume of the configuration space is $K=(2\pi)^n$. Details of the
energy functions are expected to be irrelevant for the algorithmic
questions addressed here. Our test case will be the small brain 
peptide Met-Enkephalin, which features 24 dihedral angels as dynamical 
variables, see table~\ref{tab_Met_acpt} (the conventions follow 
Ref.~\cite{Ei01}, which differs from~\cite{LiSh87}). 
Besides the $\phi,\,\psi$ angles, we keep also the 
$\omega$ angles unconstrained, which are usually restricted to 
$[\pi-\pi/9,\pi+\pi/9]$. This allows us to illustrate the RM idea 
for a particularly simple case.

\begin{table}[ht]
\caption{ Acceptance rates for dihedral angle movements. They are
accurate to about $\pm 1$ in the last digit.
\label{tab_Met_acpt} }
% \protect\label{tab_Met_acpt} }
% Results files in SMMP2/Results: T400_1/amean.r, T300_1/amean.r and
%                              T300_rm1/amean.r.
\medskip
\centering
\begin{tabular}{|c|c|c|c|c|c|} 
var     & angle  & residues & 400$\,$K& 300$\,$K& 300$\,$K
                                                            \\ \hline
        &        &        & Metro   & Metro &RM$_1$\\ \hline
$v_1$   &$\chi^1$& Tyr-1  & 0.107   & 0.070 & 0.272\\ \hline
$v_2$   &$\chi^2$& Tyr-1  & 0.182   & 0.128 & 0.343\\ \hline
$v_3$   &$\chi^6$& Tyr-1  & 0.497   & 0.377 & 0.680\\ \hline
$v_4$   &$\phi$  & Tyr-1  & 0.392   & 0.340 & 0.547\\ \hline
$v_5$   &$\psi$  & Gly-2  & 0.096   & 0.044 & 0.139\\ \hline
$v_6$   &$\omega$& Gly-2  & 0.049   & 0.034 & 0.416\\ \hline
$v_7$   &$\phi$  & Gly-2  & 0.112   & 0.045 & 0.076\\ \hline
$v_8$   &$\psi$  & Gly-3  & 0.106   & 0.038 & 0.064\\ \hline
$v_9$   &$\omega$& Gly-3  & 0.041   & 0.025 & 0.301\\ \hline
$v_{10}$&$\phi$  & Gly-3  & 0.088   & 0.035 & 0.070\\ \hline
$v_{11}$&$\psi$  & Phe-4  & 0.115   & 0.040 & 0.077\\ \hline
$v_{12}$&$\omega$& Phe-4  & 0.047   & 0.030 & 0.368\\ \hline
$v_{13}$&$\chi^1$& Phe-4  & 0.109   & 0.086 & 0.277\\ \hline
$v_{14}$&$\chi^2$& Phe-4  & 0.192   & 0.166 & 0.403\\ \hline
$v_{15}$&$\phi$  & Phe-4  & 0.082   & 0.042 & 0.139\\ \hline
$v_{16}$&$\psi$  & Met-5  & 0.122   & 0.063 & 0.156\\ \hline
$v_{17}$&$\omega$& Met-5  & 0.062   & 0.047 & 0.573\\ \hline
$v_{18}$&$\chi^1$& Met-5  & 0.117   & 0.092 & 0.362\\ \hline
$v_{19}$&$\chi^2$& Met-5  & 0.159   & 0.121 & 0.585\\ \hline
$v_{20}$&$\chi^3$& Met-5  & 0.269   & 0.211 & 0.709\\ \hline
$v_{21}$&$\chi^4$& Met-5  & 0.455   & 0.385 & 0.833\\ \hline
$v_{22}$&$\phi$  & Met-5  & 0.129   & 0.086 & 0.258\\ \hline
$v_{23}$&$\psi$  & Met-5  & 0.378   & 0.267 & 0.469\\ \hline
$v_{24}$&$\omega$& Met-5  & 0.114   & 0.096 & 0.873\\ \hline
$E$     &        &        & 0.168   & 0.119 & 0.375\\
\end{tabular} \end{table} \vspace*{0.2cm}

The Metropolis importance sampling would be perfected, if we could 
propose new configurations $\{v_i'\}$ with their canonical pd. This 
is not possible as no Metropolis simulation would be necessary if the 
canonical pd were known. But conventional Metropolis simulations 
work well at sufficiently high temperatures $T'$ and can thus provide 
an {\it estimate} ${\overline\rho}(v_1,\dots,v_n;T')$ of the pd 
$\rho(v_1,\dots,v_n;T')$. Due to the funnel picture, we expect 
that such an {\it estimate} can be used to feed useful information 
into the simulation at a sufficiently close-by lower temperature 
$T<T'$~\cite{Be03}. The idea of the  RM scheme is to propose a 
transition from a configuration $\{v_i\}$ to a new configuration 
$\{v_i'\}$ with the pd ${\overline \rho}(v'_1,\dots,v'_n;T')$ and 
to accept it with the probability
\begin{equation} \label{P0_acpt}
  P_a = \min \left[ 1, {\exp\left(-\beta\,E'\right)\,
  {\overline\rho}(v_1,\dots,v_n;T') \over \exp\left(-\beta\,E\right)
  \, {\overline\rho}(v'_1,\dots,v'_n;T')} \right]
\end{equation}
where $\beta=1/(kT)$. This equation biases the a-priori probability 
of each dihedral angle with an estimate of its pd from a higher 
temperature. Arbitrary gneralized ensembles can be treated similarly
by replacing $\exp(-\beta\,E')$ and $\exp(-\beta\,E)$ in 
Eq.~(\ref{P0_acpt}) by the appropriate probabilities $P_g(E')$ and
$P_g(E)$ of the generalized ensemble.

For a range of temperatures
\begin{equation} \label{T_order}
 T_1\ >\ T_2\ >\ \dots\ >\ T_r\ >\ \dots\ >\ T_{f-1}\ >\ T_f
\end{equation}
the simulation at the highest temperature, $T_1$, is performed 
with the usual Metropolis algorithm and the results are used as
input for the simulation at $T_2$. The estimated pd
${\overline\rho}(v_1,\dots,v_n;T_{r-1})$ is expected to be a 
useful approximation of $\rho(v_1,\dots,v_n;T_r)$, therefore alllowing
the scheme to zoom in on the native structure that is dominant at the 
physically relevant final temperature $T_f$.

To get things started, we need to construct an estimator 
${\overline\rho}(v_1,\dots,v_n;T_r)$ from the numerical data 
of the RM simulation at temperature $T_r$. Although this is 
neither simple nor straightforward, a variety of approaches offer 
themselves to define and refine the desired estimators. 

In Ref.~\cite{Be03} the approximation
\begin{equation} \label{rho0_T0}
{\overline\rho}(v_1,\dots,v_n;T_r) = \prod_{i=1}^n
{\overline\rho}^1_i(v_i;T_r) 
\end{equation}
was investigated, where ${\overline\rho}^1_i(v_i;T_r)$ are 
estimators of reduced one-variable pds defined by
\begin{equation} \label{pd1}
 \rho^1_i(v_i;T) = \int_{-\pi}^{+\pi} \prod_{j\ne i} d\,v_j\,
 \rho(v_1,\dots ,v_n;T)\ .
\end{equation}
The resulting algorithm, called RM$_1$, constitutes the simplest RM 
scheme possible. 

\begin{figure}[-t] \begin{center}
\epsfig{figure=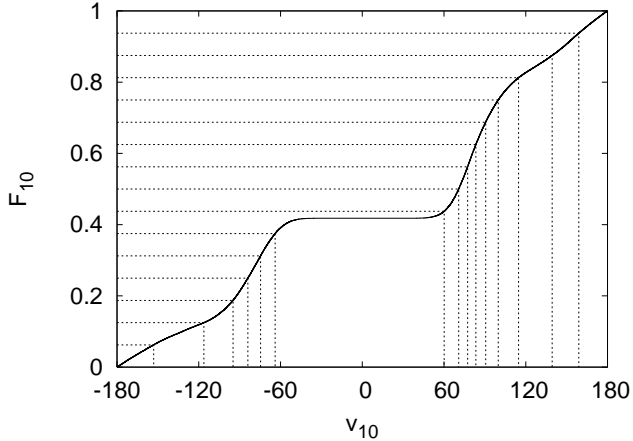,width=\columnwidth} \vspace{-1mm}
\caption{Estimate of the cumulative distribution function for the 
Met-Enkephalin dihedral angle $v_{10}$ (Gly-3 $\phi$) at $400\,K$.
\label{fig_Met_df10} }
\end{center} \vspace{-3mm} \end{figure}

Let us fill in the details of the RM$_1$ implementation~\cite{Be03}.
To update with the RM$_1$ weights it is convenient to rely on the 
cumulative distribution functions defined by
\begin{equation} \label{df1}
  F_i(v)\ =\ \int_{-\pi}^v dv'\, \rho^1_i(v')\ .
\end{equation}
The estimate of $F_{10}$, the cumulative distribution function 
for the dihedral angle Gly-3 $\phi$ ($v_{10}$), from the vacuum 
simulations at our highest temperature, $T_1=400\,$K, is shown in 
Fig.~\ref{fig_Met_df10} (this is the same angle for which histograms 
at $400\,$K and $300\,$K are shown in Ref.~\cite{Be03}). For our 
plots in the present paper we use degrees, while we use radiant in 
our theoretical discussions and in the computer programs. 
Fig.~\ref{fig_Met_df10} is obtained by sorting all 
$n_{\rm dat}$ values of $v_{10}$ in our time series in ascending 
order and increasing the values of $F_{10}$ by $1/n_{\rm dat}$ 
whenever a measured value of $v_{10}$ is encountered. Using a heapsort 
approach, the sorting is done in $n_{\rm dat}\,\log_2(n_{\rm dat})$ 
steps (see, e.g., Ref.~\cite{bbook}).

Next we divide the ordinate between 0 and 1 into $n_{\rm tab}$ equal 
segments. The value of $n_{\rm tab}$ has to be small enough that a 
table of size $n\times n_{\rm tab}$ fits conveniently into the computer 
RAM. For each integer $j=1,\dots, n_{\rm tab}$ the value $F_{i,j} = 
j/n_{\rm tab}$ defines a unique value $v_{i,j}$ through $F_{i,j} = 
F_i(v_{i,j})$ as is indicated in the figure (for which $i=10$). 
Furthermore, for each choice of a dihedral angle (i.e., a particular
value of $i$) we define the differences
\begin{equation} \label{del_vij}
  \triangle v_{i,j}=v_{i,j}-v_{i,j-1}~~{\rm with}~~v_{i,0}=-\pi\ .
\end{equation}
The grid in Fig.~\ref{fig_Met_df10} shows the discretization for 
the variable $v_{10}$ and the choice $n_{\rm tab}=16$. While the 
discretization for $F_{10}$ on the ordinate is uniformly spaced, 
widely varying intervals are obtained for $v_{10}$ on the abscissa.
The Metropolis procedure for one update of a dihedral angle $v_i$ is 
now specified as follows:

\begin{enumerate}

\item Place the present angle $v_i$ on the discretization grid, i.e.,
      find the integer $j$ through the relation $v_{i,j-1}\le v_i < 
      v_{i,j}$. For one-variable updates $j$ is available in the 
      computer memory if it is stored at the previous update. 
      Otherwise, $j$ can be re-calculated in $n_2$ steps
      for the choice $n_{\rm tab}=2^{n_2}$~\cite{BaBe05}. 

\item Pick an integer $j'$ uniformly distributed in the range 
      $1$ to $n_{\rm tab}$.

\item Propose $v'_i=v_{i,j'-1}+x^r\,\triangle v_{i,j'}$, where 
      $0\le x^r<1$ is a uniformly distributed random number.

\item Accept $v'_i$ with the probability 
      \begin{equation} \label{pa_RM1}
        p_a = \min\left[ 1, { \exp (-\beta E')\, \triangle v_{i,j'} 
        \over \exp (-\beta E)\, \triangle v_{i,j} }\right]\ .
       \end{equation}

\end{enumerate}

It is through the widely varying ratios 
$\triangle v_{i,j'} / \triangle v_{i,j}$ that importance 
sampling for the rugged variables becomes improved. Back to
our illustration in Fig.~\ref{fig_Met_df10}: The short and 
the long interval on the abscissa are proposed with equal
probabilities, i.e., the a-priori probability density for our angle
is high in short intervals and low in long intervals. The CPU
time consumption of the RM$_1$ scheme is practically identical with 
that of the conventional Metropolis algorithms, because the bulk
of the CPU time is spend on the calculation of the new energy $E'$.

\subsection{Numerical results \label{sub_RM1} }

The performance of the RM$_1$ algorithm is tested at 300$\,$K using 
input from a simulation at 400$\,$K. The temperature of 400$\,$K is 
high enough so that the conventional Metropolis algorithm is efficient, 
while it is low enough to provide useful input for the simulation at
300$\,$K, a temperature at which one experiences a considerable 
slowing down in a conventional Metropolis simulation of Met-Enkefalin.

Our Metropolis simulations are performed with a variant of SMMP 
(Simple Molecular Mechanics for Proteins)~\cite{Ei01}. For each 
simulation a time series of $2^{17}=131,072$ configurations 
is kept, sampling every 32 sweeps. A sweep is defined by updating 
each dihedral angle once, which we do in the 
sequential order of the angles listed in table~\ref{tab_Met_acpt}. 
Usually sequential updating is more efficient than random 
updating~\cite{bbook}. Before starting with the measurements,
$2^{18}=262,144$ sweeps are performed for reaching equilibrium. 
Thus, the entire simulation at one temperature uses $2^{18} +
32\cdot 2^{17}=4,456,448$ sweeps. On a $1.9\,$GHz Athlon PC this takes 
under 12 hours. For each dihedral angle the acceptance rate of the 
Metropolis algorithm was monitored at run time and, following the 
recipes of \cite{bbook}, the integrated autocorrelation time 
$\tau_{\rm int}$ is calculated from the recorded time series. 

Acceptance rates for dihedral angle movements are compiled in
table~\ref{tab_Met_acpt}. For the energy entry it is the ratio of all 
accepted over all proposed moves. Results are given for simulations 
with the conventional Metropolis algorithm at 400$\,$K and 300$\,$K, 
and for the RM$_1$ simulations at 300$\,$K. The RM$_1$ updating uses
a discretization with $n_{\rm tab}=2^7=128$ from the 400$\,$K 
Metropolis data. Acceptance rates greater than~0.3 are 
desirable~\cite{bbook}. From the table we notice that the acceptance 
rates vary greatly from angle to angle.  For the Metropolis simulation 
the values are in the interval $[0.041,0.497]$ at 400$\,$K and in 
$[0.025,0.387]$ at 300$\,$K. For both temperatures $v_9$ corresponds 
to the lowest value, while $v_3$ and $v_{21}$ correspond to the highest 
values.

Our RM$_1$ updating at 300$\,$K increases the acceptance rate for each
angle, often even beyond the Metropolis acceptance rate at 400$\,$K, as 
is obvious from the average value listed for the energy. A second look 
reveals that the increase in the acceptance rate varies greatly from 
angle to angle. While for some angles the problem of low acceptance 
rates is entirely solved, for others the improvement remains modest.
For instance for all $\omega$ angles the increase is dramatic, e.g.,
from 0.034 to 0.416 for $v_6$. Angles with little improvements are
$v_7\,(0.045\to 0.076)$, $v_8\,(0.038\to 0.064)$, $v_{10}\,(0.035\to 
0.070)$, and $v_{11}\,(0.040\to 0.077)$. Better, but still not 
particularly impressive, is the increase in the acceptance rates 
of $v_5$, $v_{15}$ and $v_{16}$. All these are $\phi,\,\psi$ angles 
around $C_{\alpha}$ atoms. For all other angles RM$_1$ updating has 
moved the acceptance rate above or at least close to~0.3.

\begin{figure}[-t] \begin{center}
\epsfig{figure=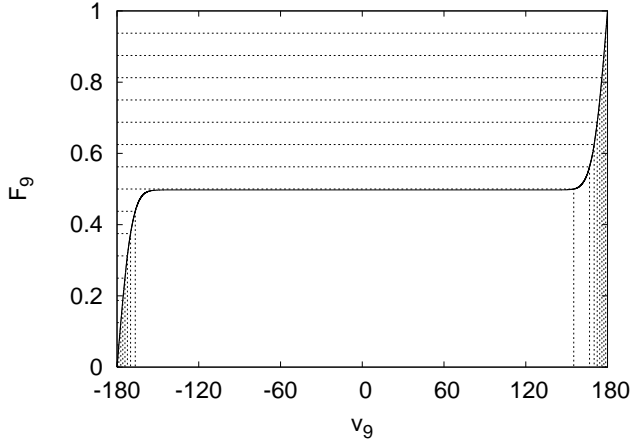,width=\columnwidth} \vspace{-1mm}
\caption{Estimate of the cumulative distribution function for the 
Met-Enkephalin dihedral angle $v_{9}$ (Gly-2 $\omega$) at $400\,$K.
\label{fig_Met_df09} }
\end{center} \vspace{-3mm} \end{figure}

The improvement for $\omega$ angles is most easily understood. 
Figure~\ref{fig_Met_df09} shows the cumulative distribution function 
for $v_9$ (Gly-2 $\omega$) at $400\,K$, which is the angle of lowest 
acceptance rate in the conventional Metropolis updating. This 
distribution function corresponds to a histogram narrowly peaked 
around $\pm \pi$, which is explained by the specific electronic 
hybridization of the CO-N peptide bond. From the grid shown in 
Fig.~\ref{fig_Met_df09} it is seen that the RM$_1$ updating 
concentrates the proposal for this angle in the range slightly 
above $-\pi$ and slightly below $+\pi$. Thus the procedure has a 
similar effect as the often used restriction to the range 
$[\pi-\pi/9,\pi+\pi/9]$, which is also the default implementation 
in SMMP (the range $[\pi,\pi+\pi/9]$ is, of course, $[-\pi,-\pi+\pi/9]$ 
in our plots).

Although acceptance rates give some insights, the decisive quantity 
for the performance of an algorithm is the more difficult to calculate
integrated autocorrelation time $\tau_{\rm int}$. To achieve a 
pre-defined accuracy, the computer time needed is directly proportional 
to $\tau_{\rm int}$. 
% \tiny
\begin{table}[ht]
\caption{Integrated autocorrelation times for dihedral angle movements
in units of 32 sweeps. \label{tab_Met_auto} }
% Results files: 
\medskip
\centering
\begin{tabular}{|c|c|c|c|c|} 
% &angle & res~\cite{Ei01}
 var    & 400$\,$K  & 300$\,$K   & 300$\,$K   & 300$\,$K \\ \hline
        & Metro     & Metro      & RM$_1$     & RM$_2$   \\ \hline
$v_1$   & 2.11 (06) &15.2 (1.5) &  9.07 (58) & 6.03 (47)\\ \hline
$v_2$   & 1.18 (02) & 2.70 (16) &  1.63 (09) & 1.70 (12)\\ \hline
$v_3$   & 1.03 (01) & 2.18 (14) &  1.26 (04) & 1.24 (04)\\ \hline
$v_4$   & 1.44 (03) & 4.44 (23) &  3.28 (21) & 2.82 (14)\\ \hline
$v_5$   & 5.44 (20) &54.5 (5.4) & 26.3 (1.5) &20.0 (1.3)\\ \hline
$v_6$   & 2.95 (07) &23.3 (2.7) &  8.65 (58) & 6.00 (34)\\ \hline
$v_7$   & 5.83 (29) &103 (14)   & 52.9 (4.3) &24.3 (1.3)\\ \hline
$v_8$   & 7.36 (22) &125 (12)   & 74.2 (6.9) &35.0 (2.7)\\ \hline
$v_9$   & 4.39 (13) &32.0 (2.2) & 14.2 (1.0) & 8.84 (48)\\ \hline
$v_{10}$& 9.08 (88) &124 (12)   & 80.6 (6.9) &34.3 (2.8)\\ \hline
$v_{11}$& 5.39 (45) &105 (08)   & 72.4 (5.5) &31.3 (1.9)\\ \hline
$v_{12}$& 3.37 (08) &15.6 (1.5) &  5.68 (39) & 3.92 (17)\\ \hline
$v_{13}$& 1.81 (05) & 8.79 (46) &  5.69 (54) & 3.59 (22)\\ \hline
$v_{14}$& 1.15 (02) & 1.65 (10) &  1.40 (07) & 1.26 (06)\\ \hline
$v_{15}$& 6.72 (28) &105 (12)   & 45.6 (2.7) &27.5 (4.5)\\ \hline
$v_{16}$& 9.28 (28) &133 (09)   & 75.2 (5.2) &33.9 (2.1)\\ \hline
$v_{17}$& 1.90 (04) & 9.69 (79) &  3.89 (36) & 2.29 (08)\\ \hline
$v_{18}$& 1.66 (05) &12.0 (1.6) &  6.48 (78) & 5.11 (28)\\ \hline
$v_{19}$& 1.17 (02) & 1.65 (08) &  1.16 (03) & 1.17 (03)\\ \hline
$v_{20}$& 1.02 (01) & 1.08 (02) &  1.03 (02) & 1.02 (02)\\ \hline
$v_{21}$& 1.00 (01) & 1.00 (01) &  1.00 (01) & 1.02 (01)\\ \hline
$v_{22}$& 3.20 (12) &35.9 (4.0) & 18.2 (1.2) &12.0 (0.8)\\ \hline
$v_{23}$& 1.50 (04) &20.3 (1.8) & 11.0 (0.6) & 5.96 (35)\\ \hline
$v_{24}$& 1.07 (02) & 1.22 (05) &  1.00 (01) & 1.00 (01)\\ \hline
 $E$    & 4.89 (21) &50.7 (5.0) & 26.0 (1.4) &14.2 (0.7)\\
\end{tabular} \end{table} \vspace*{0.2cm}

In table~\ref{tab_Met_auto} the integrated autocorrelation times 
are compiled for all angles and for the energy. The values are 
statistically consistent with those of Ref.~\cite{Be03}. Deviations
are due to re-runs and using different procedures for estimating
integrated autocorrelation times. In all tables they are given in units 
of 32 sweeps, as this is the step-size of our MC time series. Error 
bars are shown in parenthesis.  For these calculations we use the 
routines of Ref.~\cite{bbook} together with a jackknife error analysis 
as explained there. The angles $v_7,\,v_8,\,v_{10},\,v_{11},v_{15}$ and 
$v_{16}$ exhibit autocorrelation times $> 100$ in the conventional 
Metropolis simulation at $300\,$K. Note that four of these
are those with the worst improvement of acceptance rates when 
moving to the RM$_1$ updating, while the remaining two belong to 
the subsequent group with still rather poor improvement.

The increase in magnitude in the autocorrelation times for these six 
angles is remarkable when the temperature of the conventional 
Metropolis simulation is lowered from 400$\,$K to 300$\,$K. This 
shows that the standard Metropolis algorithm is efficient at 400$\,$K 
but not so at 300$\,$K. On the other hand, the distribution of the 
variables is not dramatically changed, at least
to the extent that this can be judged from one-variable histograms,
as is illustrated in Ref.~\cite{Be03} for $v_{10}$. This is the 
reason why the 400$\,$K simulation provides useful input for the
RM$_1$ simulation at 300$\,$K.

The RM$_1$ updating reduces the integrated autocorrelation times at
300$\,$K by factors of about two, for instance for $v_7$ from 103 to 
53. The $\tau_{\rm int}$ values vary greatly from angle to angle. 
While some angles show no autocorrelations after 32 sweeps 
($\tau_{\rm int}=1$ or close to it), the largest value on record for 
RM$_1$ updating at 300$\,$K is $\tau_{\rm int}=80\pm 7$ for $v_{10}$
(down from 124 for Metropolis updating at 300$\,$K). That the RM$_1$
updating does not reduce the large autocorrelation times more 
efficiently has obviously to do with correlations between different
angles. Notably even moves of some of the $\omega$ angles, like $v_9$ 
with $\tau_{\rm int}=14.2\pm 1.0$, appear considerably correlated with 
the rest of the molecule. RM variants which move several dynamical 
variables collectively are required and our RM$_2$ implementation for 
simultaneous updates of two dihedral angles is discussed in 
section~\ref{sec_RM2}. First let us address a multi-hit Metropolis 
procedure.

\subsection{Multi-hit updating}\label{sub_mhit}

Our sequential updating hits each angle once. The greatly varying 
integrated autocorrelation times of table~\ref{tab_Met_auto} suggest
that the computer time may be more efficiently used by performing 
several Metropolis hits for variables with large integrated 
autocorrelation times, to be called ``bad'' variables in the following. 

To find an optimal choice for the number of hits per variable requires 
some thought. At 300$\,$K the integrated autocorrelation times of the 
dihedral angles vary between $\tau_{\rm int}=1$ and $\tau_{\rm int}
\approx 133$ for the conventional Metropolis updating and still between
$\tau_{\rm int}=1$ and $\tau_{\rm int}\approx 80$ for the RM$_1$ 
algorithm. It is certainly not a good idea to choose the number of 
hits per variable in proportion to $\tau_{\rm int}$, because we expect 
correlations between angles to be the main obstacle for reducing large 
integrated autocorrelation times. A scheme with a large number of hits 
mimics the heat-bath algorithm (e.g., Ref.~\cite{bbook}), which 
sets the upper bound to the gain in performance, but does not resolve 
the problem of correlations between angles. So a modest increase in the 
number of hits per bad variable may increase the performance of the 
updating, while a further increase will result in the contrary.

\begin{table}[ht]
\caption{ Multi-hit performance. \label{tab_Met_mhit} }
% Results files in SMMP2/Results: 
\medskip
\centering
\begin{tabular}{|c|c|c|c|c|c|c|} 
var    & 48 & 400$\,$K & 300$\,$K & 39 & 300$\,$K & 300$\,$K \\ \hline
       &hits& Metro    & Metro    &hits&RM$_1$    & RM$_2$   \\ \hline
$v_1$   & 2 & 2.05 (07)& 14.2 (1.1)& 1 & 7.29 (52)& 5.66 (42)\\ \hline
$v_2$   & 1 & 1.37 (03)&  3.29 (23)& 1 & 1.56 (04)& 1.91 (05)\\ \hline
$v_3$   & 1 & 1.04 (04)&  2.15 (10)& 1 & 1.39 (04)& 1.51 (06)\\ \hline
$v_4$   & 1 & 1.47 (03)&  5.49 (57)& 1 & 2.74 (12)& 3.09 (16)\\ \hline
$v_5$   & 3 & 3.73 (08)& 48.1 (5.7)& 2 &21.8 (1.7)&20.3 (1.0)\\ \hline
$v_6$   & 1 & 4.19 (09)& 19.3 (1.2)& 1 & 7.12 (35)& 5.11 (22)\\ \hline
$v_7$   & 4 & 3.96 (21)& 61.7 (3.5)& 4 &42.3 (2.9)&20.7 (1.2)\\ \hline
$v_8$   & 4 & 5.06 (19)& 81.8 (5.2)& 4 &50.5 (4.1)&24.1 (1.2)\\ \hline
$v_9$   & 2 & 4.21 (13)& 25.0 (1.6)& 1 &12.7 (1.0)& 8.14 (47)\\ \hline
$v_{10}$& 4 & 5.28 (20)& 86.1 (6.4)& 4 &48.0 (4.4)&24.7 (1.6)\\ \hline
$v_{11}$& 4 & 3.72 (14)& 81.1 (7.2)& 4 &53.7 (4.8)&25.5 (1.4)\\ \hline
$v_{12}$& 2 & 3.04 (13)& 11.3 (0.6)& 1 & 4.82 (42)& 4.61 (26)\\ \hline
$v_{13}$& 1 & 2.39 (05)&  9.36 (96)& 1 & 4.36 (29)& 3.45 (26)\\ \hline
$v_{14}$& 1 & 1.34 (03)&  2.03 (15)& 1 & 1.23 (03)& 1.28 (03)\\ \hline
$v_{15}$& 4 & 4.65 (16)& 61.1 (4.4)& 2 &35.5 (3.2)&20.3 (1.0)\\ \hline
$v_{16}$& 4 & 6.29 (20)& 86.7 (8.5)& 2 &61.5 (5.0)&29.5 (1.9)\\ \hline
$v_{17}$& 1 & 2.86 (06)&  7.48 (42)& 1 & 3.16 (30)& 2.18 (07)\\ \hline
$v_{18}$& 1 & 2.03 (05)& 11.7 (1.1)& 1 & 7.79 (91)& 6.48 (51)\\ \hline
$v_{19}$& 1 & 1.64 (04)&  2.57 (15)& 1 & 1.16 (02)& 1.32 (03)\\ \hline
$v_{20}$& 1 & 1.09 (01)&  1.21 (02)& 1 & 1.01 (01)& 1.02 (01)\\ \hline
$v_{21}$& 1 & 1.00 (01)&  1.02 (02)& 1 & 1.00 (01)& 1.00 (01)\\ \hline
$v_{22}$& 2 & 2.98 (08)& 26.2 (1.8)& 1 &19.2 (1.6)&13.9 (0.9)\\ \hline
$v_{23}$& 1 & 1.51 (05)& 13.3 (1.1)& 1 &11.4 (1.0)& 5.60 (31)\\ \hline
$v_{24}$& 1 & 1.44 (03)&  1.63 (04)& 1 & 1.02 (01)& 1.03 (02)\\ \hline
$E$     &   & 4.25 (17)& 32.9 (1.4)&   &24.9 (2.2)& 14.7 (1.3)\\
\end{tabular} \end{table} \vspace*{0.2cm}

A guideline for choosing the number of hits is obtained from the 
observation that the previously obtained acceptance rates per update 
attempt do not change when performing multiple hits. It appears 
reasonable to increase the hits of bad variables while bounding the 
number of hits times the acceptance rate by 0.3 from above. As the 
acceptance rates change considerably when switching from regular
Metropolis to RM$_1$ updating, we employ different schemes for the two 
cases. Results for the two different multi-hit schemes are collected 
in table~\ref{tab_Met_mhit}. 

The numbers in the first ``hits'' column are used for the 
regular Metropolis updating. They are arranged to add up to 48, i.e., 
twice the total number of variables. The additional computer time 
needed is balanced by reducing the number of sweeps between 
measurements from 32 to 16 (a sweep is now defined by applying 
the new updating procedure in sequential order once to each angle). 
By comparing tables~\ref{tab_Met_auto} and \ref{tab_Met_mhit} we see 
that the multi-hit updating improves the Metropolis algorithm at 
300$\,$K considerably: the integrated autocorrelation time for the 
energy is down by about 40\%.

The numbers in the second ``hits'' column are used for RM$_1$ and 
RM$_2$ updating. As RM$_1$ updating increases acceptance rates already
significantly, there is little opportunity for additional improvements
due to multiple hits. By that reason the numbers of the column add only
up to 39 hits per sweep. This is balanced by reducing the number of
sweeps between measurements from 32 to 20 (the integer nearest to
$32\times 24/39$). There are still significant decreases in    
autocorrelations times for the bad variables, but the indicator 
for overall performance, the integrated autocorrelation
time of the energy, shows only a modest 5\% decrease 
when comparing to RM$_1$ without multiple hits and practially no 
change for RM$_2$ updating, introduced next. The apparent 
reason is that these updating schemes are already much closer to a 
heat-bath scenario, so that the improvement due to multiple hits 
becomes offset by the additional computer time needed.

\section{The RM$_2$ approximation}\label{sec_RM2}

We now generalize the RM$_1$ scheme
of Eq.~(\ref{pa_RM1}) to the simultaneous updating of two dihedral 
angles. For $i_1\ne i_2$ the reduced two-variable pds are defined by
\begin{equation} \label{pd2}
  \rho^2_{i_1,i_2}(v_{i_1},v_{i_2};T) = \int_{-\pi}^{+\pi} 
  \prod_{j\ne i_1,i_2} d\,v_j\, \rho(v_j,\dots ,v_n;T)\ .
\end{equation}
The one-variable cumulative distribution functions $F_{i_1}$ and the 
discretization $v_{i_1,j},\,j=0,\dots,n_{\rm tab}$ are already
given by Eqs.~(\ref{df1}) and~(\ref{del_vij}). We define conditional
cumulative distribution functions by
\begin{equation} \label{df2}
  F_{i_1,i_2;j}(v) = \int_{-\pi}^v dv_{i_2}
  \int_{v_{i_1,j-1}}^{v_{i_1,j}} dv_{i_1}\,
  \rho^2_{i_1,i_2}(v_{i_1},v_{i_2})
\end{equation}
for which the normalization $F_{i_1,i_2;j}(\pi)=1/n_{\rm tab}$ holds.
To extend the RM$_1$ updating to two variables we define for each 
integer $k=1,\dots,n_{\rm tab}$ the value $F_{i_1,i_2;j,k} =
k/(n_{\rm tab})^2$. Next we define $v_{i_1,i_2;j,k}$ through 
$F_{i_1,i_2;j,k} = F_{i_1,i_2;j}(v_{i_1,i_2;j,k})$ and also the 
differences
\begin{equation} \label{del_vkj}
  \triangle v_{i_1,i_2;j,k} = v_{i_1,i_2;j,k} - v_{i_1,i_2;j,k-1}  
  ~~{\rm with}~~ v_{i_1,i_2;j,0} = -\pi\ .
\end{equation}
The RM$_2$ Metropolis procedure for the simultaneous update of 
$(v_{i_1},v_{i_2})$ is then specified as follows:

\begin{enumerate}
\item Find the grid index $j$ for the present angle $v_{i_1}$ through
      $v_{i_1,j-1}\le v_{i_1}\le v_{i_1,j}$, just like for RM$_1$ 
      updating.
\item Find the grid index $k$ for the present angle $v_{i_2}$ through
      $v_{i_1,i_2;j,k-1}\le v_{i_2}\le v_{i_1,i_2;j,k}$.
\item Pick two integers $j'$ and $k'$, each uniformly distributed in 
      the range $1$ to $n_{\rm tab}$. (This could be extended to cover 
      asymmetric ranges $n^1_{\rm tab}\times n^2_{\rm tab}$.)
\item Propose $v'_{i_1}=v_{i_1,j'-1}+x^r_1\,\triangle v_{i_1,j'}$, 
      where $0\le x^r_1<1$ is a uniformly distributed random number.
\item Propose $v'_{i_2}=v_{i_1,i_2;j',k'-1}+x^r_2\,\triangle 
      v_{i_1,i_2;j',k'}$, where $0\le x^r_2<1$ is a second uniformly 
      distributed random number.
\item Accept $(v'_{i_1},v'_{i_2})$ with the probability 
\begin{equation} \label{pa_RM2}
  p_a^2 = \min \left[ 1 , { \exp (-\beta E')\, \triangle v_{i_1,j'}\, 
  \triangle v_{i_1,i_2;j',k'}\over \exp (-\beta E)\,\triangle v_{i_1,j}
  \, \triangle v_{i_1,i_2;j,k}\, }\right]\ .
\end{equation}
\end{enumerate}

As before, estimates of the conditional cumulative distribution 
functions and the intervals $\triangle v_{i_1,i_2;j,k}$ are 
obtained from the conventional Metropolis simulation at 400$\,$K. In 
the following we focus on the pairs $(v_7,v_8)$, $(v_{10},v_{11})$ 
and $(v_{15},v_{16})$. These angles correspond to the largest 
integrated autocorrelation times of the RM$_1$ procedure and are 
expected to be strongly correlated with one another because they are 
pairs of dihedral angles around a $C_{\alpha}$ atom.

\begin{figure}[-t] \begin{center}
% Created with SMMP2/Results/T400_1/dia*.plt.
\epsfig{figure=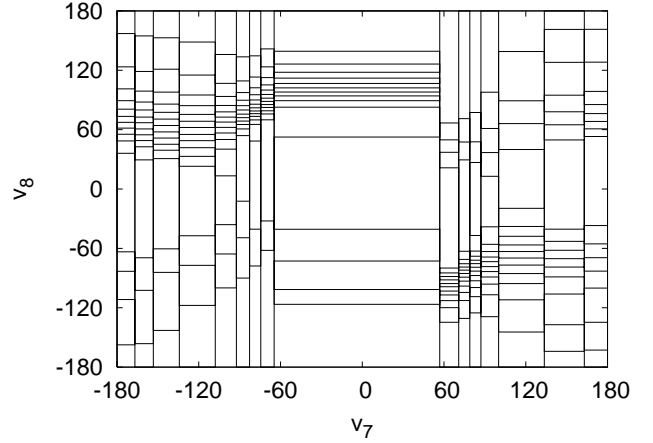,width=\columnwidth} \vspace{-1mm}
\caption{Areas of equal probabilities (sorting $v_7$ then $v_8$).
\label{fig_d07_d08}}
\end{center} \vspace{-3mm} \end{figure}

\begin{figure}[-t] \begin{center}
% Created with SMMP2/Results/T400_1/dia*.plt.
\epsfig{figure=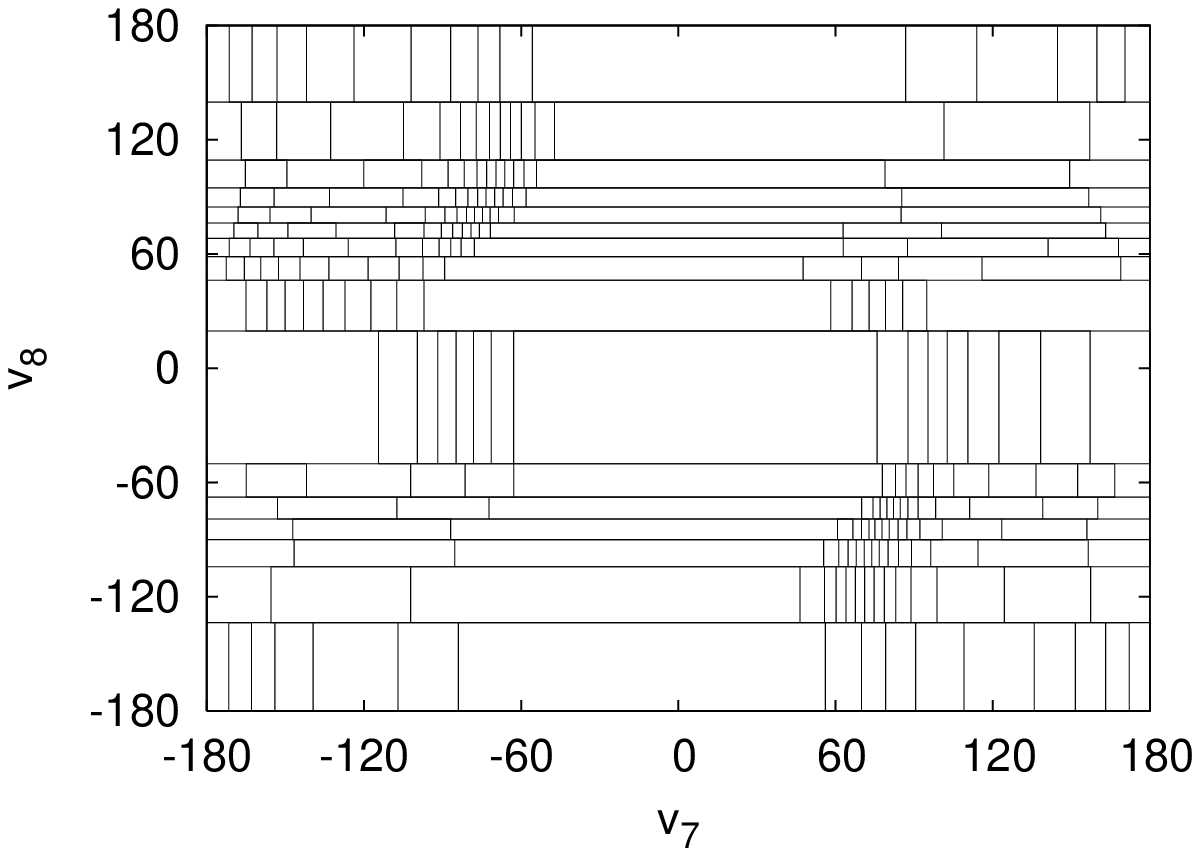,width=\columnwidth} \vspace{-1mm}
\caption{Areas of equal probabilities (sorting $v_8$ then $v_7$).
\label{fig_d08_d07}}
\end{center} \vspace{-3mm} \end{figure}

\begin{figure}[-t] \begin{center}
% Created with SMMP2/Results/T400_1/dia*.plt.
\epsfig{figure=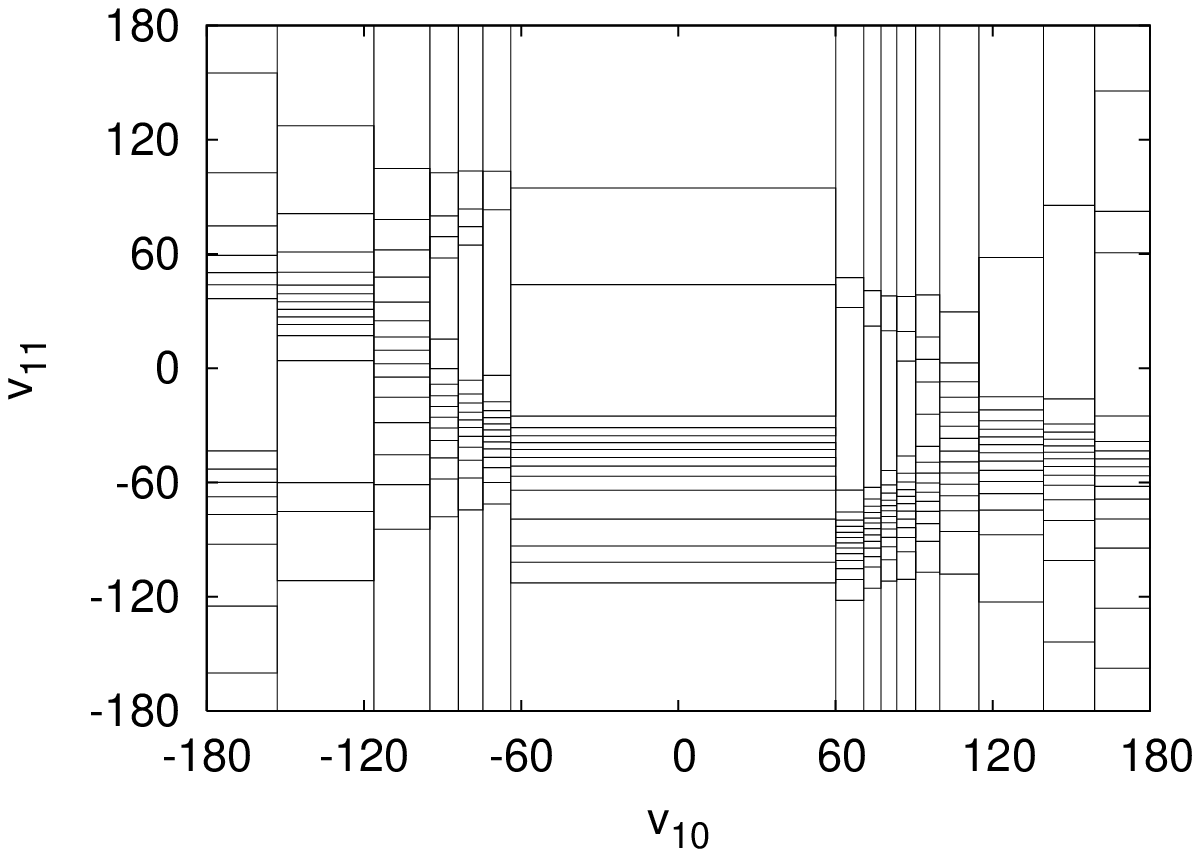,width=\columnwidth} \vspace{-1mm}
\caption{Areas of equal probabilities (sorting $v_{10}$ then $v_{11}$).
\label{fig_d10_d11}}
\end{center} \vspace{-3mm} \end{figure}

\begin{figure}[-t] \begin{center}
% Created with SMMP2/Results/T400_1/dia*.plt.
\epsfig{figure=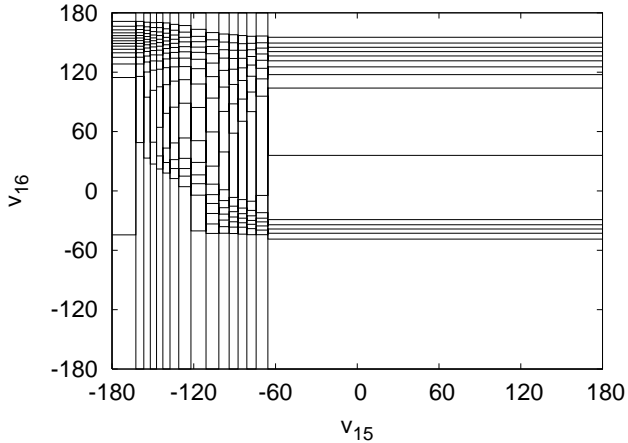,width=\columnwidth} \vspace{-1mm}
\caption{Areas of equal probabilities (sorting $v_{15}$ then $v_{16}$).
\label{fig_d15_d16}}
\end{center} \vspace{-3mm} \end{figure}

The bias of the acceptance probability given in Eq.~(\ref{pa_RM2}) is 
governed by the areas 
$$ \triangle A_{i_1,i_2;j,k} = \triangle v_{i_1,j}\, 
   \triangle v_{i_1,i_2;j,k}\ . $$
For $i_1=6$ and $i_2=7$ our 400$\,$K estimates of these areas are
depicted in Fig.~\ref{fig_d07_d08}. For the RM$_2$ procedure these
areas take the role which the intervals on the abscissa of 
Fig.~\ref{fig_Met_df10} play for RM$_1$ updating. The small and 
the large areas are proposed with equal probabilities, so the 
a-priori probability for our two angles is high in a small area
and low in a large area. In Fig.~\ref{fig_d07_d08} the largest 
area is 503.4 times the smallest area. Areas of high probability
correspond to allowed regions in the Ramachandran map of a Gly
residue~\cite{Sch79}.

Note that the order of the angles matters. The difference between 
Fig.~\ref{fig_d07_d08} and Fig.~\ref{fig_d08_d07} is that we plot 
in Fig.~\ref{fig_d07_d08} the areas $A_{7,8;j,k}$ and in 
Fig.~\ref{fig_d08_d07} the areas $A_{8,7;j,k}$ while the labeling 
of the axes is identical. This means that for Fig.~\ref{fig_d07_d08} 
sorting is first done on the angle $v_7$ (regardless of the value 
of $v_8$) and then done on $v_8$ for which the corresponding value 
of $v_7$ is within a particular bin $\triangle v_7$, but for 
Fig.~\ref{fig_d08_d07} it is first done one $v_8$ and then on $v_7$. 
In Fig.~\ref{fig_d08_d07} the largest area is 396.4 times the smallest 
area.

Fig.~\ref{fig_d10_d11} and Fig.~\ref{fig_d15_d16} give plots for 
the $(v_{10},v_{11})$ and $(v_{15},v_{16})$ pairs in which the angle 
with the smaller subscript is sorted first. The ratio of the largest 
area over the smallest area is 650.9 for $(v_{10},v_{11})$ 
and 2565.8 for $(v_{15},v_{16})$. The large number in the latter 
case is related to the fact that $(v_{15},v_{16})$ is the pair of
$\phi,\,\psi$ angles around the $C_{\alpha}$ atom of Phe-4, for
which positive $\phi$ values are disallowed \cite{Sch79}.

The RM$_2$ scheme which we have tested adds updates for the three pairs 
$(v_7,v_8)$, $(v_{10},v_{11})$ and $(v_{15},v_{16})$ after one-angle
updates for all the 24 angles with the RM$_1$ scheme. For each pair 
both orders of sorting are used, so that we add altogether six new 
updates. The bookkeeping for this process is a bit tricky, because an 
accepted update changes not only $(j,k)\to (j',k')$, but also the the 
$j$ from the RM$_1$ updating of the angles. The latter corresponds
to a different table and needs to be re-calculated from the new 
value of the angle. As already mentioned, this can be done in 
$\log_2(n_{\rm tab})$ steps~\cite{BaBe05}. Similarly, accepted RM$_1$ 
updates can change the initial RM$_2$ $(j,k)$ values, so that they 
may have to be re-calculated. 
The six RM$_2$ update tables, each of size $16\times 16$, are built 
from the 400$\,$K Metropolis simulation, and the areas of four of 
them are precisely those shown in Figs.~\ref{fig_d07_d08} 
to~\ref{fig_d15_d16}.  

\subsection{Numerical Results}\label{sec_NumRM2}

We have checked the correctness of our 
updating procedure by comparing high precision energy averages and 
other observables with results from previous calculations.  The
acceptance rates of the one-variable updates remain the same as 
they were for RM$_1$ procedure. For the acceptance rate of a pair 
we average over the two cases. Table~\ref{tab_RM2_acpt} compares 
the two-angle RM$_2$ acceptance rates at 300$\,$K to those obtained 
by proposing the same two-angle updates with the standard Metropolis 
procedure. At 300$\,$K an increase by factors in the range from 
three to nearly ten is found. However, the values remain surprisingly 
low, presumably due to substantial correlations with additional angles.
\begin{table}[ht]
\caption{ Acceptance rates for simultaneous moves of angle pairs.
\label{tab_RM2_acpt} }
% Results files in SMMP2/Results: 
\medskip \centering
\begin{tabular}{|c|c|c|c|c|} 
variable pair    & 400$\,$K & 300$\,$K & 300$\,$K& 300$\,$K\\ \hline
                 & Metro    & Metro    & RM$_2$  & RM$_2$  \\ \hline
                 & & &$n_{\rm tab}=16$ &$n_{\rm tab}=128$ \\ \hline
$(v_7,v_8)$      & 0.044    & 0.0060   & 0.019   & 0.020  \\ \hline
$(v_{10},v_{11})$& 0.041    & 0.0051   & 0.021   & 0.022  \\ \hline
$(v_{15},v_{16})$& 0.018    & 0.0051   & 0.048   & 0.050  \\
\end{tabular} \end{table} \vspace*{0.2cm}

Integrated autocorrelation times are calculated to evaluate the
improvement of the overall performance. For this purpose the number 
of sweeps between measurements is reduced from 32 to 26 to account for 
the additional CPU time needed for the two-angle moves. The results 
are presented in table~\ref{tab_Met_auto}. Despite the small 
acceptance rates for the two-angle moves, the integrated 
autocorrelation times for the targeted angles are substantially 
reduced. For all the six angles they are smaller by factors larger than 
two when compared with the RM$_1$ results. Interestingly this speed-up 
propagates through the entire system and the integrated autocorrelation 
time for the energy is found to be about a factor of two smaller 
than for the RM$_1$ algorithm.

Multi-hit updates allow us to focus even more on the angles with large
autocorrelation times. For the one-angle updates we use the same 
numbers of hits as for RM$_1$ updating (see table~\ref{tab_Met_mhit}). 
In addition we use four hits for the pairs $(v_7,v_8)$ and 
$(v_{10},v_{11})$ and two hits for the $(v_{15},v_{16})$ pair. 
For each each pair both orders of the updating are used. Altogether 
we perform 39 one-angle and 20 two-angle hits per 
sweep, which is balanced by reducing the number of sweeps between 
measurements to 13 (the integer nearest to $32\times 24/59$). The 
results for integrated autocorrelation times are a mixed bag
(compare the last columns of tables~\ref{tab_Met_auto} 
and~\ref{tab_Met_mhit}). While the values for the targetted 
angles indeed go down, in particular for $v_8$, the improvement 
does not propagate to the energy. 

\begin{table}[ht]
\caption{ Multi-hit Metropolis at low
          temperatures. \label{tab_Met_lowT} }
% Results files in SMMP2/Results: 
\medskip
\centering
\begin{tabular}{|c|c|c|c|c|c|} 
var     & 48 & 280$\,$K  &  260$\,$K  &  240$\,$K   & 220$\,$K    \\ \hline
        &hits&           &            &             &             \\ \hline
$v_1$   & 2 &  24.5 (1.7)&  59.8 (4.8)&  246 (18)   &  658 (60)   \\ \hline
$v_5$   & 3 &  70.7 (5.0)& 236 (27)   &  531 (47)   & 1672 (148)  \\ \hline
$v_6$   & 1 &  34.8 (2.4)&  65.0 (4.8)&  185 (10)   &  453 (28)   \\ \hline
$v_7$   & 4 & 156 (16)   & 526 (88)   & 1425 (118)  & 4434 (405)  \\ \hline
$v_8$   & 4 & 191 (17)   & 627 (78)   & 1591 (135)  & 4965 (485)  \\ \hline
$v_9$   & 2 &  60.2 (5.0)& 114 (06)   &  467 (33)   & 1809 (219)  \\ \hline
$v_{10}$& 4 & 186 (15)   & 613 (69)   & 1530 (131)  & 5326 (443)  \\ \hline
$v_{11}$& 4 & 214 (18)   & 625 (82)   & 1901 (148)  & 5781 (437)  \\ \hline
$v_{12}$& 2 &  20.1 (1.1)&  44.3 (3.2)&  158 (08)   &  557 (44)   \\ \hline
$v_{13}$& 1 &  15.7 (05) &  36.8 (2.9)&  119 (07)   &  279 (17)   \\ \hline
$v_{15}$& 4 & 239 (90)   & 308 (31)   &  999 (78)   & 2283 (170)  \\ \hline
$v_{16}$& 4 & 204 (12)   & 449 (34)   & 1600 (129)  & 4537 (344)  \\ \hline
$v_{17}$& 1 &  11.9 (0.7)&  22.8 (1.2)&   78.8 (2.7)&  278 (15)   \\ \hline
$v_{18}$& 1 &  24.1 (2.6)&  55.5 (3.6)&  208 (11)   &  755 (46)   \\ \hline
$v_{22}$& 2 &  64.5 (4.7)& 136 (10)   &  343 (16)   & 898 (54)    \\ \hline
$v_{23}$& 1 &  37.2 (1.8)&  84.4 (5.1)&  360 (26)   & 830 (63)    \\ \hline
$E$     &   &  71.6 (4.0)& 148 (08)   &  484 (29)   & 1536 (121)  \\
\end{tabular} \end{table} \vspace*{0.2cm}

\begin{table}[ht]
\caption{ RM$_2$ multi-hit at low temperatures. \label{tab_RM2_lowT} }
% Results files in SMMP2/Results: 
\medskip
\centering
\begin{tabular}{|c|c|c|c|c|c|} 
var     & 39 & 280$\,$K &  260$\,$K  &  240$\,$K  & 220$\,$K    \\ \hline
        &hits&          &            &            &             \\ \hline
$v_1$   & 1 & 11.6 (0.7)&  27.5 (1.2)&  81.6 (6.2)&  188 (017)  \\ \hline
$v_5$   & 2 & 36.9 (1.8)&  89.5 (7.4)& 256 (26)   &  824 (103)  \\ \hline
$v_6$   & 1 & 10.7 (0.6)&  20.1 (0.7)&  50.4 (2.4)&  146 (017)  \\ \hline
$v_7$   & 4 & 45.0 (3.1)& 105 (05)   & 359 (27)   & 1434 (164)  \\ \hline
$v_8$   & 4 & 63.0 (4.0)& 131 (07)   & 459 (42)   & 2266 (309)  \\ \hline
$v_9$   & 1 & 17.5 (0.8)&  44.9 (1.9)& 130 (10)   &  405 (030)  \\ \hline
$v_{10}$& 4 & 59.9 (4.1)& 134 (09)   & 514 (57)   & 2344 (300)  \\ \hline
$v_{11}$& 4 & 57.1 (3.6)& 157 (10)   & 495 (35)   & 1965 (178)  \\ \hline
$v_{12}$& 1 &  8.04 (30)&  20.1 (1.0)&  52.9 (3.4)&  131 (008)  \\ \hline
$v_{13}$& 1 &  6.55 (25)&  12.9 (0.7)&  39.0 (3.5)&   86.7 (6.8)\\ \hline
$v_{15}$& 2 & 43.5 (6.4)&  88.8 (4.9)& 265 (18)   &  775 (082)  \\ \hline
$v_{16}$& 2 & 58.7 (2.8)& 140 (08)   & 420 (32)   & 1569 (164)  \\ \hline
$v_{17}$& 1 &  4.43 (20)&   9.72 (44)&  30.7 (1.9)&   69.2 (4.4)\\ \hline
$v_{18}$& 1 & 10.7 (0.5)&  26.5 (1.2)&  65.6 (2.6)&  206 (12)   \\ \hline
$v_{22}$& 1 & 25.4 (1.3)&  53.6 (2.7)& 124 (05)   &  281 (014)  \\ \hline
$v_{23}$& 1 & 11.6 (0.6)&  28.2 (1.7)&  88.2 (5.2)&  267 (022)  \\ \hline
$E$     &   & 22.9 (1.1)&  47.9 (1.8)& 128 (09)   &  426 (038) \\
\end{tabular} \end{table} \vspace*{0.2cm}

In tables~\ref{tab_Met_lowT} and~\ref{tab_RM2_lowT} we present results 
down to 220$\,$K for the multi-hit improved Metropolis and RM$_2$ 
algorithms, where the distribution of hits is the same as listed
in table~\ref{tab_Met_mhit}. We always construct the RM$_2$ table 
from the previous RM$_2$ runs at the next higher temperature. 
We also generate RM$_2$ data without using the multi-hit scheme,
with the resulting autocorrelation times consistenly higher 
than those reported in table~\ref{tab_RM2_lowT}. We have not listed
the angles $v_2,\,v_3,\,v_4,\,v_{14},\, v_{19},\, v_{20},\,v_{21}$ 
and $v_{24}$ in tables~\ref{tab_Met_lowT} and~\ref{tab_RM2_lowT}, 
because they are not significantly correlated with the rest of the 
molecule.

When lowering the temperature towards 220$\,$K, the autocorrelation 
times increase rapidly. To control $\tau_{\rm int}$
we double the number of sweeps between measurements each time, when
decreasing the temperature. So it is 64 at 280$\,$K, 128 at 260$\,$K,
256 at 240$\,$K and 512 at 220$\,$K. For the multi-hit Metropolis
simulations this was still not sufficient and we performed two
additional runs, with $4\times 256$ sweeps at 240$\,$K and with 
$4\times 512$ sweeps at 220$\,$K. This explains the relatively small
error bars in the last two columns of table~\ref{tab_Met_lowT}. In the 
tables we continue to report $\tau_{\rm int}$ in units of 32 sweeps,
multiplying the measured $\tau_{\rm int}/32$ value with the number of 
sweeps between measurements. It is seen that even for the RM$_2$ 
simulation the $\tau_{\rm int}$ increases are not compensated by
the increase of computer time. In contrast to that, the decrease in 
acceptance rates is rather moderate, less than a factor of two 
when the temperature is lowered from $300\,$K to $220\,$K.

The results of tables~\ref{tab_Met_lowT} and~\ref{tab_RM2_lowT} show 
that our RM$_2$ sampling accelerates the conventional Metropolis 
simulations by a rather temperature independent factor. As we can
assume that the multi-hit Metropolis simulations already improve
conventional Metropolis simulations by about 40\%, the RM$_2$
accelaration is by a factor between four and five with respect 
to a conventional Metropolis simulation. For large scale simulations 
factors larger than two are clearly of importance, but it remains a 
bit puzzling why the improvement does not increase upon lowering 
the temperature, as it is found when using generalized ensembles. 
Apparently coordinated moves of three and more angles are needed. 

\section{Summary and conclusions}\label{sec_summary}

We have reviewed the one-variable approximation RM$_1$ of the 
rugged Metropolis (RM) scheme of Ref.~\cite{Be03} and worked out
a two-variable approximation RM$_2$ for simultaneous moves of two 
dihedral angles. As before the test system has been Met-Enkephalin.
A gain of a factor of four over conventional Metropolis simulations 
has been demonstrated at 300$\,$K. 

Although the elaboration of the RM scheme seems to be on 
track, much work is left to be done. Even for a system as simple as
Met-Enkephalin it remains unclear which kinds of correlations are 
responsible for the still low acceptance rates of the two-angles moves. 
On the other hand it is encouraging to see that the autocorrelations 
times of these angles are nevertheless substantially reduced and that 
this effect propagates through the entire system. Other test cases 
need to be investigated to get a broader understanding of the 
observed features. In particular one would like to know how the
performance gain depends on the system size.

Somewhat puzzling is the lack of enhanced improvements at lower 
temperatures. The real future of biased updating procedures may 
lie in their implementation for generalized ensembles.

Presently the leading method for simulations of biomolecules is 
molecular dynamics (see \cite{FrSm96} for a textbook). This is to some 
extent surprising, because Markov Chain Monte Carlo (MC) simulations 
allow for large changes of conformations in a single move, so 
thermodynamically relevant equilibrium configurations can, in 
principle, be reached quickly. However, in simulations of biomolecules 
with an explicit inclusion of solvent interactions, large MC moves face 
the problem that there will not be a suitable cavity in the solvent to 
accommodate a large distortion of the molecule shape. While the RM 
method discussed in this paper decreases the likelihood of steric 
clashes in a vacuum simulation, it has no immediate translation into 
the situation of explicit solvent models. 

The way out may be the use of implicit solvent models, for which the 
change in the molecule-solvent and solvent-solvent interaction energies 
can be calculated instantaneously, like in a vacuum simulation. Indeed 
RM$_1$ simulations for implicit solvent models, based on the 
solvent-accessible area method implemented in \cite{Ei01}, have
already been performed~\cite{BH04}. The algorithmic improvements 
were similar as found for the vacuum situation. However there is 
evidence \cite{BH04,PeHa03} that the class of solvent models used
does not parametrize the solvent interactions properly. It appears 
that quite generally the reliability of implicit solvent 
models has not yet been well established. 

Finally we like to mention that MC moves may be fine-tuned on a local 
level as done in the approach of Ref.~\cite{Swendsen}. This is also 
possible for models which include solvents explicitly. So MC may 
still be a viable alternative to molecular dynamics for explicit
solvent models.

\acknowledgments
Our calculations were performed on the Anfinsen PC cluster of 
FSU's School of Computational Science. H.-X. Zhou was supported
in part by the National Institutes of Health grant GM~58187.

\clearpage
\end{document}